%% file: ShdHO_arXiv.tex
\documentclass[
amsmath,amssymb,
aps,
prl, reprint,
]{revtex4-1}

\usepackage{amssymb,bm}
\usepackage{graphicx}

\usepackage[%
colorlinks=true,
urlcolor=blue,
linkcolor=blue,
citecolor=blue
]{hyperref} 
\usepackage{amsmath}
\usepackage{epstopdf}
\usepackage{float}
\allowdisplaybreaks

\usepackage[usenames]{color}

\begin{document}
	
	\title{\textbf{Spin mixing between subbands and extraordinary Landau levels shift in wide HgTe quantum wells}}
	
	\author{A.A.~Dobretsova$^{*\,1,2}$, A.D.~Chepelianskii$^{\,3}$, N.N.~Mikhailov$^{\,1,2}$ and Z.D.~Kvon$^{\,1,2}$}
	\affiliation{$^1$Rzhanov Institute of Semiconductor Physics, Novosibirsk 630090, Russia}
	\affiliation{$^2$Novosibirsk State University, Novosibirsk 630090, Russia}
	\affiliation{$^3$LPS, Universite Paris-Sud, CNRS, UMR 8502, F-91405 Orsay, France}

	\date{July 05, 2018}
	
	\begin{abstract}
		We present both the experimental and theoretical investigation of a non-trivial electron Landau levels shift in magnetic field in wide $\sim$20\,nm HgTe quantum wells: Landau levels split under magnetic fields but become degenerate again when magnetic field increases. We reproduced this behavior qualitatively within an isotropic 6-band Kane model, then using semiclassical calculations we showed this behavior is due to the mixing of the conduction band with total spin 3/2 with the next well subband with spin 1/2 which reduces the average vertical spin from 3/2 to around 1. This change of the average spin changes the Berry phase explaining the evolution of Landau levels under magnetic field.
	\end{abstract}	
	
	
	\maketitle

	\begin{figure}[H]
		\begin{minipage}[h]{1\linewidth}
			\begin{tabular}{p{0.3\linewidth}p{0.6\linewidth}}
				(a) & (b)
			\end{tabular}
		\end{minipage}
		\begin{minipage}[h]{0.3\linewidth}
			\flushright\includegraphics[scale=0.25]{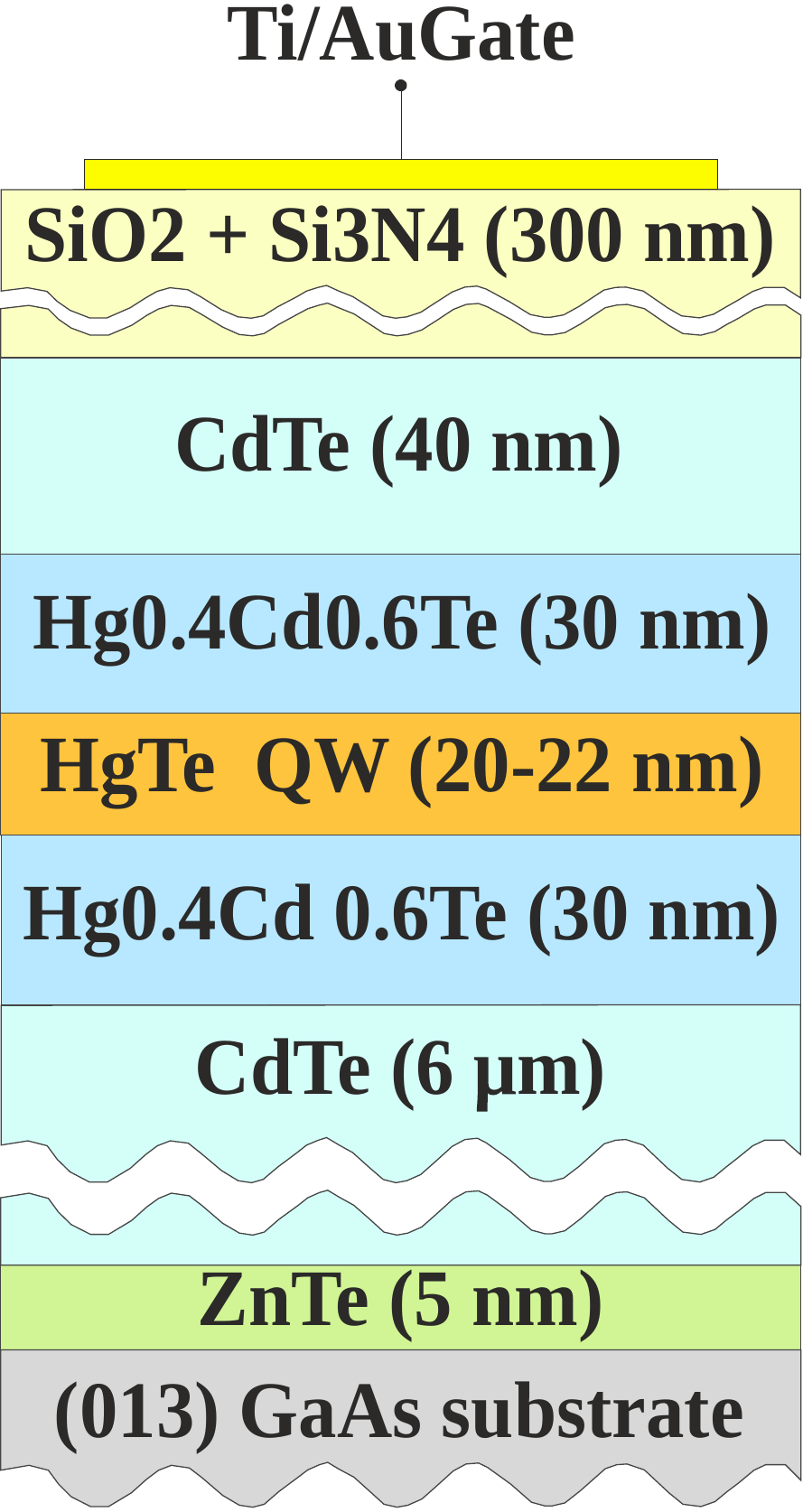}
		\end{minipage}
		\begin{minipage}[h]{0.67\linewidth}
			\flushright\includegraphics[scale=0.8]{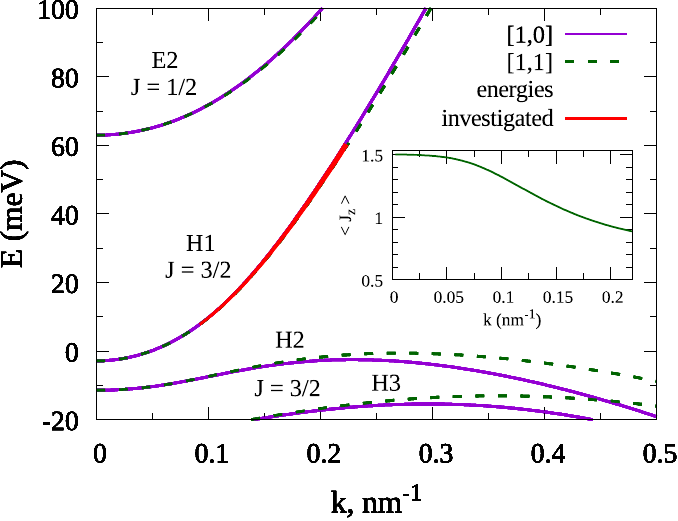}
		\end{minipage}
		\caption{(a)~The cross section of the heterostructures studied. (b)~Energy spectrum of (001) 20\,nm HgTe strained quantum well calculated by 6-band Kane model. Solid lines correspond to $\textbf{k}$ direction [1,~0], dashed lines - [1,~1]. \textit{Inset} - an average $z$-component of the conduction band total angular momentum $<J_z>$ in an isotropic model as a function of the wave vector module~$k$.}
		\label{pic:sample}
	\end{figure}
	
	\section{Introduction}
	
	The recent discovery of topological insulators, a new state of mater, in materials with strong spin-orbit interaction~\cite{Kane2005, Bernevig2006, Konig2007, Fu2007, Murakami2007, Hsieh2008} has opened an exciting research direction. In most topological insulators the strong relativistic effects and particularly spin orbit coupling lead to an inverted spectrum which is at the origin of their topological properties. Since relativistic effects strength grows with increasing atomic mass, investigations have focused on a broad range of materials with Bi and Sb~\cite{Ando2013,Chen2009,Hsieh2009,Zhang2011,Taskin2011}, InAs/(In)GaSb~\cite{Knez2011,Knez2014,Akiho2016} and HgTe~\cite{Konig2007,Roth2009, Gusev2011} quantum wells, all containing elements with a large atomic numbers in Mendeleev periodic table. The quantum wells based on HgTe and InAs\,/\,GaSb can be grown epitaxially and the carrier density can be tuned with a gate in a broad range, these systems are thus particularly suited for fundamental investigations on topological insulators. One of the principal feature of HgTe quantum wells is that their properties drastically depend on their thickness. By changing the width~$d$ of the quantum well~\cite{Raichev2012a} one can get: an ordinary insulator like GaAs quantum wells at $d < 6.3$\,nm~\cite{Bernevig2006a,Konig2007,Roth2009}; a linear spectrum of so called Dirac fermions at the critical thickness $d\approx6.3$\,nm~\cite{Buttner2011a,KvonDF2011}; inverted spectrum of 2D topological insulator at $d > 6.3$\,nm~\cite{Bernevig2006a,Konig2007,Roth2009,Gusev2011}; 2D semimetallic spectrum with a small overlap between the conduction and valence bands at $d\approx20$\,nm~\cite{Kvon2008SM1,Kvon2011rev}; the spectrum of 3D topological insulator at width $d~>~70$\,nm~\cite{Brune2011,Crauste2013,Kozlov2014}. 
	
	In addition to its impact on the band structure, spin orbit interaction can appear directly in magneto-transport experiments through the shift of Landau levels (LLs). Previous works \cite{Nitta1997,Zhang2001,Gui2004} focused on Landau levels shift due to the Rashba splitting arising as a consequence of the asymmetric deformation of the quantum well at high gate voltages. In this work we present magneto-transport experiments and a thorough theoretical analysis showing that the shift of Landau levels can also occur due to mixing of quantum well subbands with different spin. Indeed the average spin of the quantum-well wavefunction is directly connected to Berry phase since both describe the transformational properties of the wavefunction under plane rotation. A change of the wavefunction spin thus directly leads to an anomalous shift of Landau levels due to the contribution of Berry phase in the semiclassical quantization rule for Landau levels  ~\cite{Roth1966SC1,Berry1984,Berry1989,Littlejohn1991a,Chang1995, Fuchs2010,Wright2013,Alexandradinata2017,Gao2017}. Following these theoretical works we develop a quantitative semi-classical model which reproduces the Landau levels obtained numerically in a realistic 6-band $\bold{k \cdot p}$ theory calculation, these theoretical results are then compared to the experiment.
	
	\section{Experiment}
	
	Our experiments were carried out on $20 \div 22$\,nm undoped Cd$_{0.6}$Hg$_{0.4}$Te/HgTe/Cd$_{0.6}$Hg$_{0.4}$Te heterostructures with (013) surface orientation, grown by molecular beam epitaxy. The cross section of the structure is illustrated in Fig.~\ref{pic:sample}\,(a). The detailed description of samples preparation can be found in \cite{Mikhailov2006a,Kvon2011rev}. These quantum wells have a separation of 65meV between $p$-like well conduction subband [H1] with total spin (angular momentum) 3/2 and the first excited $s$-like subband [E2] with spin 1/2. This splitting is sufficiently large to resolve a large number of Landau levels in the conduction band at few Tesla magnetic fields but still within the range covered by changes in the gate potential, $20 \div 22$nm wide quantum wells were thus optimal for our experiments. The evolution of the conduction band spin as a function of momentum $k$ is shown in the inset in Fig.~\ref{pic:sample}\,(b), the average spin in $z$ direction indeed changes from $3/2$ at low energies to around $1$ near the bottom of the subband [E2] as a consequence of their mixing. The previous works on similar HgTe samples ~\cite{Kvon2008SM1,Olshanetsky2009,Gusev2010, Kozlov2011SM, Gusev2012, Raichev2012, Minkov2013} were mostly concentrated on the energy region when the conduction and valence bands overlap (see Fig.~\ref{pic:sample}\,(b) leading to coexistence of electrons and holes. It has been shown in InAs/GaSb quantum wells~\cite{Karalic2016} that mixing between $p$-like electrons and $s$-like holes states can already lead to a non-trivial LLs behavior. In our case both the first conduction and valence subbands are $p$-like, in the work we thus concentrated on the energy region highlighted in red in Fig.~\ref{pic:sample}\,(b), which correspond to a metal regime where the two dimensional electron gas is formed only by electrons of the conduction band [H1]. To perform magnetotransport measurements, the structures were patterned into Hall bars  with width 50\,$\mu$m and the distance between the contacts 100 and 250\,$\mu$m. The structures had also a metallic top gate allowing to reach the metallic regime and to change electron densities from ${10^{10}}$\,cm$^{-2}$ to ${8\cdot10^{11}}$\,cm$^{-2}$. Electron mobility in this electron concentration range varied from 5 to 30\,m$^2$/Vs and measurements were performed by standard lock-in technique at temperature ${T = 0.2\div2}$\,K and magnetic field up to 9\,T.
	
	\begin{figure*}
		\begin{minipage}[h]{1\linewidth}
			\begin{tabular}{p{0.53\linewidth}p{0.46\linewidth}}
				(a) & (b) \\
			\end{tabular}
		\end{minipage}
		\begin{minipage}[h]{0.53\linewidth}
			\flushleft\includegraphics[scale=1.3]{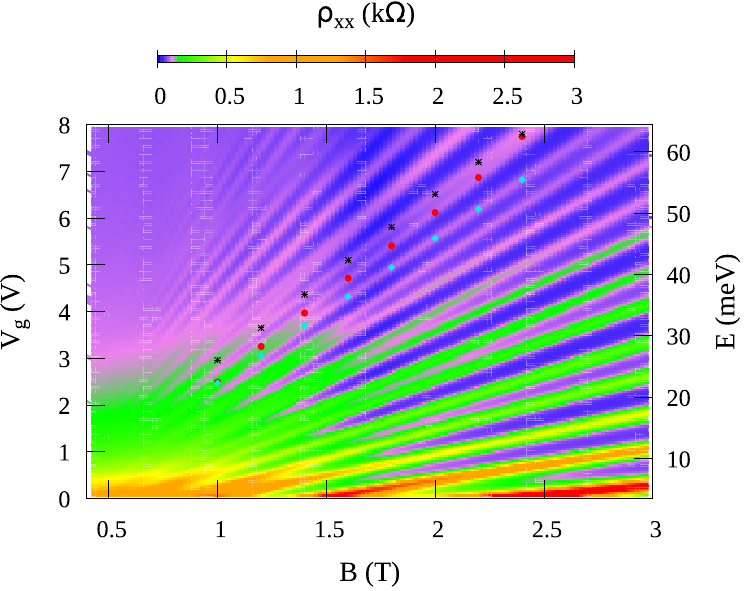}
		\end{minipage}
		\hfill
		\begin{minipage}[h]{0.46\linewidth}
			\flushleft\includegraphics[scale=1.]{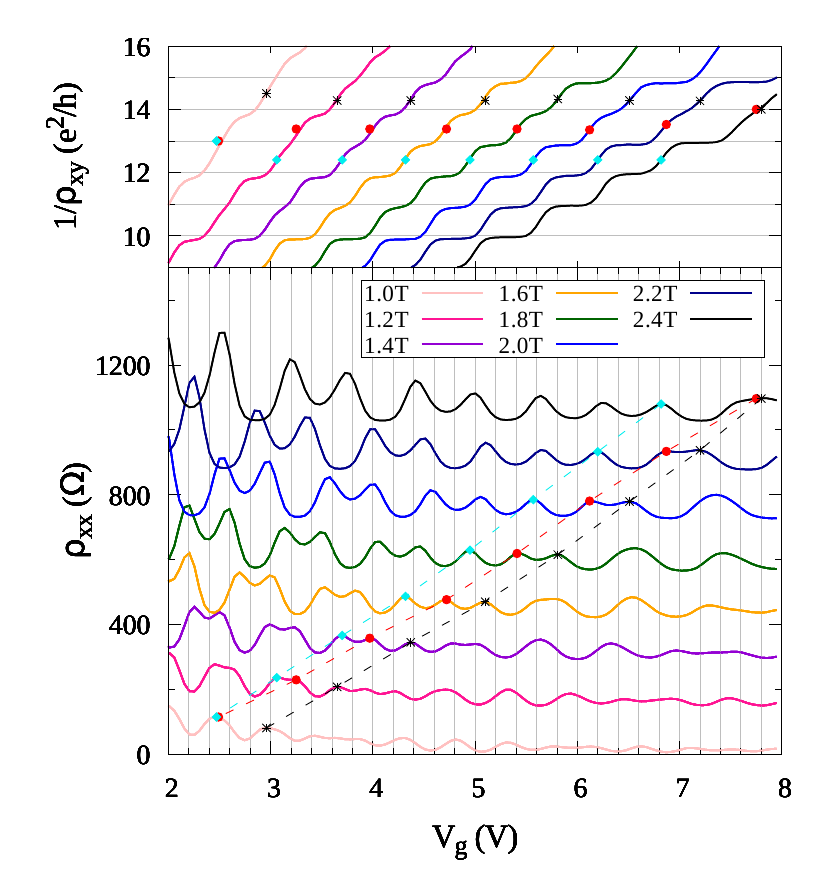}
		\end{minipage}
		\caption{(a)~2D color map of longitudinal resistivity $\rho_{xx}$ versus magnetic field $B$ and gate voltage $V_g$ for 22\,nm undoped HgTe quantum well. The energy axis calculated from $V_g$ by using 6-band Kane model is also showed. Dots highlight the example of Landau levels with non-trivial behavior: ''red'' (highlighted by red dots) and ''black'' levels being split at $B=1$\,T degenerate  at $B=2.4$\,T. (b)~Longitudinal resistivity $\rho_{xx}$ and Hall resistance $\rho_{xy}$ versus $V_g$ at magnetic field from 1.0 to 2.4\,T. Dots correspond to the same Landau levels as in 2D color map\,(a). In the $\rho_{xx}$ dependence on $V_g$ $y$-axis corresponds to $B=1.0$\,T, the each next curve is shifted up by 200\,$\Omega$.}
		\label{pic:expres}
	\end{figure*}
	
	The typical experimental results are shown in Fig.~\ref{pic:expres}, where as example results for 22\,nm HgTe quantum well are presented. In Fig.~2\,(a) 2D color map of longitudinal resistivity $\rho_{xx}$ versus magnetic field $B$ and gate voltage $V_g$ is shown. Here the energy axis calculated from $V_g$ by using 6-band Kane model is also present. In the figure one can see that some non-degenerate neighbor Landau levels (resistivity maxima) coincide with each other with magnetic field increase. One of the example of such LLs behavior is highlighted by dots: ''red'' level (highlighted by red dots) at small finite magnetic field $B \sim 1$\,T is degenerate with ''blue'' and separated from ''black'' ones however with $B$ increase it comes closer to the ''black'' one till their degeneration at $B=2.4$ while the 'blue'' level become separate.
	
	The same pattern can be also seen in Fig.~\ref{pic:expres}\,(b), where the fixed magnetic field cross-sections of $\rho_{xx}$ from Fig.~2~(a) and Hall resistance $\rho_{xy}$ are shown for $B$ from 1.0 to 2.4\,T. Here Landau levels highlighted by dots in Fig.~\ref{pic:expres}\,(a) are also highlighted by dots of the same color. In the dependence of longitudinal resistivity on gate voltage the described effect is seen as the transformation of two peaks   at $B=1$\,T: one corresponding to the ''blue'' and ''red'' levels and one separated to ''black'' one, into two different peaks at $B=2.4$\,T:  one corresponding to ''red'' and ''black'' levels and now ''blue'' one is separated. To simplify the visualization we have joined the dots of the same color. In Hall resistance dependence on gate voltage one can also see that at $B=1$\,T there is Hall plateau corresponding to filling factor 14 while at 2.4\,T it disappears. Similar results were obtained in~\cite{Minkov2016a}, but the microscopic explanation in term of Berry phase change due to spin mixing between subbands was not presented.
	
	\section{Theory}
	
\begin{figure}[t]
		\centering
		\includegraphics[scale = 1]{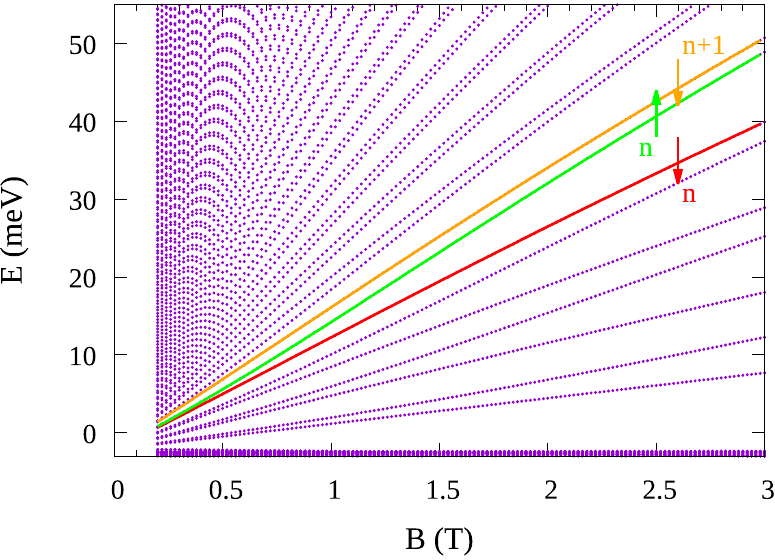}
		\caption{Landau levels position as a function of the perpendicular magnetic field in (001) 20\,nm HgTe quantum well calculated by 6-band Kane model in isotropic approximation. Orange, green and red lines show the example of Landau levels with non-trivial behavior as in the experiment: orange and green levels come closer to each other with magnetic field increase. $n$~is orbital Landau level number, arrows symbolize level spin directions.}
		\label{pic:speccalc}
	\end{figure}

	To describe the unusual Landau levels behavior observed in the experiment we have calculated the energy spectrum of (001) 20\,nm HgTe quantum well with infinite wall boundaries in magnetic field by means of 6-band Kane model as in \cite{Raichev2012a} with Kane parameters taken from  \cite{Novik2005} (see Fig.~\ref{pic:sample}\,(b)). In order to calculate the position of Landau levels as function of the perpendicular magnetic field, we used an isotropic model for the quantum well. This assumption enabled us to find eigenstates using only a small number of nearby oscillator wavefunctions and thus to solve the problem without diagonalizing very large matrices,  we checked by full diagonalization that the anisotropic terms deform the spectrum only weakly. Our numerical results for the Landau levels of the first conduction band versus magnetic field are shown in Fig.~\ref{pic:speccalc}. The calculation gives the energy levels (highlighted by yellow, green and red lines) with the same behavior as was observed in the experiment: yellow and green levels non-degenerate at small magnetic field come closer to each other at large~$B$. Note that in this calculation we didn't take into account the well asymmetry arising as a consequence of the applied gate voltage, the calculated Landau levels shift are thus an intrinsic property of the quantum well spectrum. The inclusion of an asymmetric contribution is discussed in the Appendix~D of the manuscript and doesn't change the qualitative picture.
	
	The physical origin of this unusual shift of the Landau levels can be understood from the semiclassical quantization rules. In semiclassical approximation Landau level energy can be found from the well-known Lifshitz-Onsager equation \cite{Kittel1963} $A(E_n) =(n + \gamma) 2\pi e B/c$, where $A$ is an area of a cyclotron orbit cross-section in $k$-space (constant-energy and $k_z = const$ surfaces intersection, $z$ - the direction of magnetic field $\textbf{B}$); $n$ - orbital Landau level number; $\gamma$ - some phase from 0 to 1. In the case of a free electron $\gamma$ is simply equal to 1/2, while in general case in solids:
	\begin{align*}
	& \gamma = {1\over 2} + \gamma_{s} = {1\over 2} - {1 \over 2\pi}\Gamma(C) + {1 \over 2\pi}\Gamma_{NG}.
	\end{align*}
	We denote by $\gamma_s$ the deviation of $\gamma$ from the free electron value, we will show that it is connected to the average spin of electrons. This quantity has several contributions, the first term is the Berry phase $\Gamma(C) = i \int_C \langle u_{\textbf{k}} | {\partial u_{\textbf{k}} \over \partial \textbf{k}} \rangle d \textbf{k} $ - written as a function of the envelope of the electron wave functions across the well  $|u_{\textbf{k}}\rangle$ as function of the in plane momentum $\textbf{k}$ (the semiclassical analysis uses wavefunctions without magnetic field). The second term $\Gamma_{NG}$ is a phase related to the electron orbital magnetic moment~\cite{Roth1966SC1,Berry1984,Berry1989,Littlejohn1991a,Chang1995, Fuchs2010,Alexandradinata2017,Gao2017}. We decided to designate $\Gamma_{NG}$ as ''non-geometric'' as in~\cite{Littlejohn1991a}, since it depends not only on the shape of the semiclassical trajectory in phase space but also  on the rate at which the orbit is traversed. For completeness we provide an elementary derivation of the general expression for $\Gamma_{NG}$ in the Appendix C, while here only its expression for the case of an isotropic spectrum will be discussed. 
	
	In the isotropic approximation, the envelope wavefunction $|u(k, \theta)\rangle$ as function of the momentum amplitude $k$ and of its polar angle $\theta$ can be expressed as function of $|u(k,\theta=0)\rangle$ by application of the total angular momentum operator $\hat J_z$ which describes rotations along $z$ direction perpendicular to the two dimensional electron gas: $|u(k,\theta)\rangle = e^{-3i \theta/2} e^{-i \theta {\hat J}_z} |u(k, 0)\rangle$. The phase factor $e^{-3i \theta/2}$ is introduced to make the wavefunctions single valued upon a full rotation $\theta \rightarrow \theta + 2 \pi$, it is necessary since $J_z$ is a direct sum of spin 1/2 and 3/2 operators. This relation can be formally established from the identity $e^{-i \theta {\hat J}_z} \hat H(\mathbf{k}) e^{i \theta {\hat J}_z} = \hat H( R_z(\theta)(\mathbf{k}) )$ where $R_z(\theta)$ is the rotation operator around $z$-axis by an angle $\theta$. Using this expression for $|u(k, \theta)\rangle$, we can connect the Berry phase accumulated along a cyclotron trajectory in magnetic field as a function of the average spin $\langle J_z\rangle$ in $z$ direction: 
	\begin{align}
	\Gamma(C) &= i\int d \theta \langle u(k, \theta) | \partial_\theta | u(k, \theta) \rangle = 2 \pi \langle J_z\rangle + 3\pi.  \label{eq:BPh}
	\end{align}
	We see that for an isotropic quantum well the Berry phase is directly connected to the spin projection along $z$ direction. 
	
	In isotropic approximation the non-geometric phase can be simplified to 
	\begin{align}
	\Gamma_{NG} &=  \frac{2 \pi}{d \epsilon/d k} \; {\rm Im} \langle \partial_\theta u | {\hat H}(k,\theta)- \epsilon(k) | \partial_k u  \rangle ,
	\label{eq:GNG}
	\end{align}
	where $\hat H(k, \theta)$ is the Hamiltonian of the quantum well as a function of the polar coordinates of the momentum $\mathbf{k}$ and $\epsilon(k)$ the dispersion relation of the conduction band. Due to the isotropy of the quantum well this expression is actually independent on $\theta$ and was numerically evaluated at $\theta = 0$.
	
	For a symmetric quantum well (without Rashba) spin up and spin down carriers have the same energy, making the absolute value of the spin projection in the conduction band a well defined function of energy with $|J_z|$ changing from 1.5 to 0.9 in the explored energy range (see inset of Fig.~\ref{pic:sample}\,(b)). Both Berry and the non-geometric phases change sign for opposite spins. To make $\gamma_s$ a well defined function of energy we take its positive branch for the spin-up orientation keeping in mind that the sign is reversed for the spin-down branch, we also take the values of all phases in the $(0, 2\pi)$ interval. For the following discussion it is convenient to introduce the dimensionless ratios $\Delta_{n \uparrow}$ between Zeeman splitting of Landau levels with the same orbital index $n$ and opposite spin and the orbital spacing between nearby Landau levels with spin down orientation. Indeed keeping in mind the symmetry of $\gamma_s$ and the semiclassical quantization rule these ratios can directly be expressed as function of $\gamma_s$: 
	\begin{align}
	\Delta_{n \uparrow} = {(E_{n \uparrow} - E_{n \downarrow}) \over (E_{(n+1) \downarrow} - E_{n \downarrow})} = 2 \gamma_s, \label{eq:dE}
	\end{align}
	we remind that $n$ is the orbital Landau level index. Equation (\ref{eq:dE}) allows us to compare the Landau-levels obtained in the experiment and in our 6-band Kane model simulations with the semiclassical values for $\gamma_s$ from Eqs.~(\ref{eq:BPh},\ref{eq:GNG}). For spin down we define
	$\Delta_{n \downarrow} = (E_{n \downarrow} - E_{n\uparrow})/(E_{(n-1) \uparrow} - E_{n\uparrow})$, with this choice both ratios  $\Delta_{n \uparrow}$ and  $\Delta_{n \downarrow}$ collapse semi-classically on the same dependence which simplifies future discussions. 
	
	\begin{figure*}
		\begin{minipage}[H]{1\linewidth}
			\begin{tabular}{p{0.49\linewidth}p{0.49\linewidth}}
				(a) & (b) \\
			\end{tabular}
		\end{minipage}
		\begin{minipage}[H]{0.49\linewidth}
			\center{ \includegraphics[scale = 1.1]{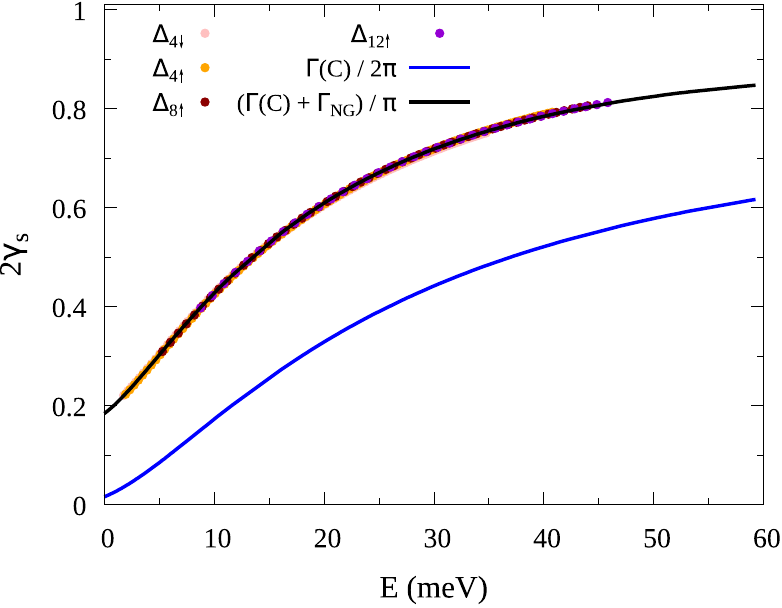}} \\
		\end{minipage}
		\hfill
		\begin{minipage}[H]{0.49\linewidth}
			\center{\includegraphics[scale = 1.1]{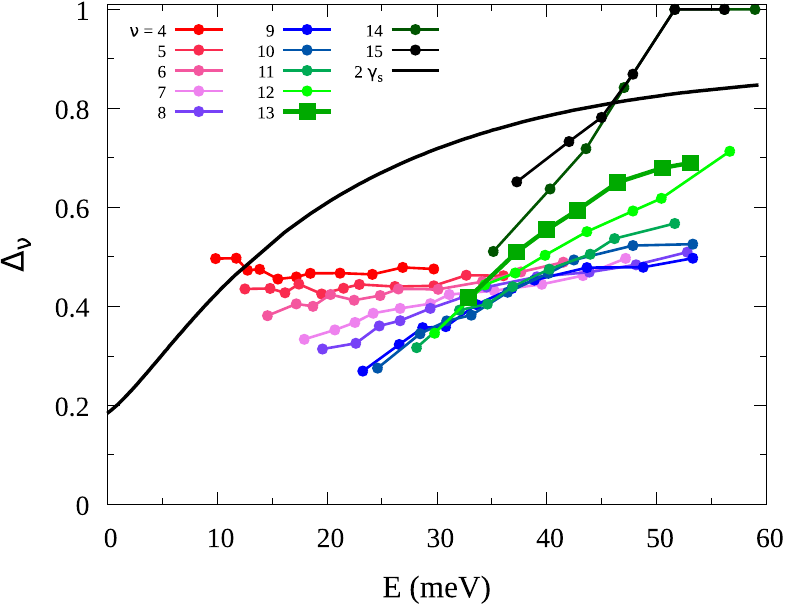}} \\
		\end{minipage}
		\caption{Comparison between the doubled deviation $2 \gamma_s$ of electron phase from free electron value 1/2 and the dimensionless ratios $\Delta_\nu$ in theory and experiment. (a) Dots show $\Delta_{n \uparrow / \downarrow}$ from 6-band Kane model in isotropic approximation; blue line - Berry phase itself (not doubled) $\Gamma(C) = 2\pi <J_z>+3\pi/2$ and the black line is $2 \gamma_s$ - the sum of Berry phase $\Gamma_{C}$ and non-geometrical phase $\Gamma_{NG}$. (b) Shows the values of $\Delta_\nu$ obtained from the experiment compared to theoretical $2\gamma_s$ (black line). Thick green line with square symbols gives $\Delta_\nu$ for the Landau level highlighted by red dots in Fig.~\ref{pic:expres}.}
		\label{pic:BP}
	\end{figure*}
	
	\section{Discussion}
	
	Fig.~\ref{pic:BP}\,(a) compares the theoretical semi-classical dependence $2 \gamma_s(E)$ with the ratios $\Delta_{n \uparrow}$ and  $\Delta_{n \downarrow}$ obtained for Landau levels from numerical diagonalisation of the isotropic 6-band Kane model - these results show an excellent agreement between the semi-classical theory and numerical Landau levels already for $n \ge 4$.  This figure also shows that the energy dependence of the total phase $\Gamma(C) + \Gamma_{NG}$ qualitatively follows the Berry phase and is approximately equal to $\Gamma(C)/2 + {\rm const}$. This approximate relation can viewed as a generalization to topological quantum wells of the exact cancellation (up to a topological winding number) between the two phases which occurs for electron-hole symmetric two band model~\cite{Fuchs2010,Wright2013}. Since for HgTe this compensation is only approximate, the anomalous shift of the Landau levels still follows the Berry phase providing a magneto-transport probe for the vertical spin projection of the quantum well wavefunctions. 
	
	The experimental data for $\Delta_\nu = \Delta_{n \uparrow,\downarrow}$ are presented in Fig.~\ref{pic:BP}\,(b) for different filling factors $\nu$, the position of the energy levels was extracted from the position of $\rho_{xx}$ maxima in Fig.~2\,(a). It is seen that while the 6-band Kane model provides a good qualitative description of the experimental energy shifts (for e.g. in Fig.~\ref{pic:speccalc}) the agreement is not quantitative. The agreement improves at high $\nu \simeq 15$ however for small $\nu \le 6$, $\Delta_{\nu}$ is independent on energy. It is possible that in the low filling factor regime transport across incompressible stripes becomes important and make transport more sensitive to density gradients at the edges of the sample~\cite{Dahlem2010}. This would make our assumption to extract the bulk energy levels position from $\rho_{xx}$ maxima invalid. On the theory side inter-Landau level exchange interactions~\cite{Ando1974,Tsitsishvili1998} that were not included in the Kane model can maybe also give a substantial $\nu$ dependent correction to the energy shifts.
	
	In summary we have observed non-trivial Landau levels shifts with magnetic field in wide $\sim$20\,nm HgTe quantum wells where Landau levels split in magnetic field and become degenerate again when magnetic field increases. Theoretical calculations within a 6-band Kane model allowed us to reproduce this picture qualitatively. We proposed a microscopical explanation based on the semiclassical quantization of the Landau levels:  this shift is caused by a mixing of states from two quantum well subbands with different spin: the conduction band with spin 3/2 and the second well subband with spin 1/2. This mixing results in a change of the average electron spin with energy, and thus a change of the Berry phase which is directly connected to the spin in an isotropic quantum well. Taking into account the Berry phase and non-geometrical phase related to the electron orbital magnetization, we reproduced theoretically the ratio between the Zeeman and orbital Landau level splittings obtained from the 6-band Kane model calculation. A quantitative agreement with the semi-classical theory could not be obtained for the energy levels extracted from our magnetotransport experiments especially at low filling factors. Further experiments on 20nm HgTe quantum wells and other topological materials are needed to understand the origin of this discrepancy.
	
	\section{Acknowledgements}
	
	We thank O.E.\,Raichev for his assistance in the implementation of the 6-band Kane model and S.\,Gueron and J.N.\,Fuchs for helpful discussions. This work was supported by the Russian Science Foundation (Grant No. 16-12-10041) and ANR grant SPINEX.
	\color{black}
	
	\input{ShdHO_arXiv_bbl.bbl}

	\clearpage
	\onecolumngrid
	
	\section{Appendix A: Hamiltonian of the system}
	
	We used the 6-band Kane model for HgTe whose matrix $\hat H(\mathbf{k})$ form can be found in \cite{Novik2005,Raichev2012a}, the matrix form however does not explicitly show that the Hamiltonian is rotation invariant in the case when $\gamma_2 = \gamma_3 = 0.9$ (). Rotation invariance can be seen from the identity $e^{-i \theta {\hat J}_z} \hat H(\mathbf{k}) e^{i \theta {\hat J}_z} = \hat H( R_z(\theta)(\mathbf{k}) )$, where $R_z(\theta)$ is the rotation operator around $z$-axis by an angle $\theta$, but it requires tedious calculations to be established.
	
	We have thus found it useful to write the HgTe Hamiltonian as function of spin operators which make rotation invariance manifest. For this purpose we introduced spin 1/2 operators $\mathbf{{\hat S}}$ and spin 1 operators $\mathbf{{\hat L}}$, their sum $\mathbf{{\hat J}} = \mathbf{{\hat S}} + \mathbf{{\hat L}}$ can then describe both spin 3/2 and 1/2 bands from the 6-band Kane model
	
	As a function of  $\mathbf{{\hat S}}$ and $\mathbf{{\hat L}}$ operators the Hamiltonian (in the isotropic approximation, and here without strain for simplisity) then reads: 
	\begin{align}
	{\hat H} = \epsilon_0(k^2) - J(k^2) \mathbf{{\hat S}} \cdot \mathbf{{\hat L}} + D \left[ \mathbf{k} \cdot (\mathbf{{\hat S}}+\mathbf{{\hat L}}) \right]^2 + \frac{2 P}{3 \sqrt{3}} \mathbf{k} \cdot \left[ \mathbf{{\hat L}} - 4 \mathbf{{\hat S}} 
	+ \mathbf{{\hat L}} \left(\mathbf{{\hat S}} \cdot \mathbf{{\hat L}} \right) + \left(\mathbf{{\hat S}} \cdot \mathbf{{\hat L}} \right) \mathbf{{\hat L}} 
	\right],
	\label{Hspin}
	\end{align}
	
	where
	\begin{align}
	\epsilon_0(k^2) &= \frac{1}{3} \left[ E_g + B k^2 ( \gamma_0 - 2 \gamma_{12}- \frac{\gamma_{23}}{2} ) \right]; \\
	J(k^2) &= \frac{2}{3} \left[ E_g + B k^2 ( \gamma_0 + \gamma_{12}- \frac{\gamma_{23}}{2} )  \right]; \\
	&D = 2 B \gamma_{23};
	\end{align}
	$E_g, ~ P, ~ B, ~ \gamma_0, ~ \gamma_{12} = (\gamma_1 + \gamma_2) \;/\; 2, ~ \gamma_{23} = (\gamma_2 + \gamma_3) \;/\; 2$ - Kane parameters of the system taken from~\cite{Novik2005}.
	
	The rotation invariance of the Hamiltonian in the form Eq.~(\ref{Hspin}) is manifest as all the terms are writen as scalar products of vectors. Most of the terms have a familiar form in the point of view of spin physics: $\epsilon_0(k^2)$ is a global energy offset, $J(k^2)$ is an exchange interaction and $D$ is a fine structure term describing an anisotropy in the $\mathbf{k}$ direction. The unusual part (from the spin-point of view) comes from the last ``Zeeman'' term proportional to $P$ which has contribution in the form of an exchange energy dependent g-factor $\left(\mathbf{k} \cdot \mathbf{{\hat L}} \right) \left(\mathbf{{\hat S}} \cdot \mathbf{{\hat L}} \right) + c.c.$. It seems likely that those terms are created by the underlying second order perturbation theory expansion in the $\bold{k \cdot p}$ theory.
	
	\section{Appendix B: Semiclassical quantization of the multicomponent Schroedinger equation}
	
	Semi-classical quantization rules have been derived in several previous publications, however we haven't found an elementary derivation in the WKB spirit that would directly apply to the multi-component wave equations that stem from $\bold{k \cdot p}$ theory. For example the derivation in \cite{Littlejohn1991a} uses the Wigner transforms formalism, this makes the derivation conceptually appealing, but also less transparent. We have chosen an approach which directly mirrors the standard derivation of the WKB approximation for the usual Sch\"oedinger equation. This allows us to make the derivation self-contained, and we have chosen to present it here.
	
	Let us assume the multicomponent 1D Schroedinger equation in the following form:
	\begin{align}
	{\hat H} &= {\hat V}_0(x) + {\hat V}_1 (-i \hbar \partial_x) + {\hat V}_2 (-i \hbar \partial_x)^2.
	\label{eq:H}
	\end{align}
	Here all the operators $V_i$ are self-adjoint; also for simplicity to avoid problems due to non-commutation between ${\hat p}_x = -i \hbar \partial_x$ and ${\hat V}_i$ we have chosen ${\hat V}_{1,2}$ to be independent on $x$. 
	
	For the next calculation it is useful to introduce:
	\begin{align*}
	{\hat H}^s(x, p_x) &= {\hat V}_0(x) + {\hat V}_1 p_x + {\hat V}_2 p_x^2.
	\end{align*}
	
	Assuming the wave function amplitude change much slower than the phase, we seek the solutions in the form:
	\begin{align}
	|\psi(x)\rangle = |u(x)\rangle e^{i S(x)/\hbar}.
	\label{eq:psi}
	\end{align}
	
	Substituting~\eqref{eq:psi} into~\eqref{eq:H} and expanding it into powers of $\hbar$, to lowest order in $\hbar$ (since in this case derivatives act only on the exponent $e^{i S(x)/\hbar}$) we have:
	\begin{align*}
	{\hat H} |\psi(x)\rangle &= {\hat H}^s(x, p_x) |u_0(x)\rangle e^{i S_0(x)/\hbar} + O(\hbar) \\
	&= E |u_0(x)\rangle e^{i S_0(x)/\hbar},
	\end{align*}
	where $p_x = \partial_x S_0$. We find that $|u_0(x)\rangle$ is an eigenvector of ${\hat H(x, p_x)}$; $x$ and $p_x$ follow Hamilton equations with an Hamiltonian given by the corresponding eigenvalue ${\hat H}^s(x, p_x)|u_0(x)\rangle = \epsilon(x, p_x) |u_0(x)\rangle$.
	
	We would like to go to higher order and seek the solution in the form $|\psi\rangle = {\cal A} |u_0(x)\rangle e^{i S_0(x)/\hbar} + \hbar |u_1(x)\rangle e^{i S_0(x)/\hbar}$ ($|u_0(x)\rangle$ and $|u_1(x)\rangle$ are normalized to unity, $\langle u_1(x) | u_0(x)\rangle = 0$).
	
	\textit{Note: we denote by triangular brackets $\langle|\rangle$ the matrix multiplying only, without integration.}
	
	For further calculations it is useful to write the following identities:
	\begin{align}
	& e^{-i S_0(x)/\hbar} (-i \hbar \partial_x) \left[ {\cal A}(x) e^{i S_0(x)/\hbar}  |u_0(x)\rangle  \right] = S_0'(x) {\cal A}(x) |u_0(x)\rangle - i \hbar \partial_x \left[ {\cal A}(x) |u_0(x)\rangle  \right] \nonumber\\
	&  e^{-i S_0(x)/\hbar} (-i \hbar \partial_x)^2 \left[ {\cal A}(x) e^{i S_0(x)/\hbar}  |u_0(x)\rangle  \right] = \label{eq:ident}\\
	&\;\;\;~~~~~~~~~~~~~~~~~ {\cal A}(x) |u_0(x)\rangle \left[ S_0'(x)^2 - i \hbar S''_0(x) \right]  - 2 i \hbar \partial_x  \left[ {\cal A}(x) |u_0(x)\rangle  \right] S_0'(x) + O(\hbar^2) \nonumber.
	\end{align}
	
	Now using~\eqref{eq:ident} we compute the matrix element:
	\begin{align*}
	&e^{-i S_0/\hbar} \langle u_0 | {\hat H} |\psi(x)\rangle = e^{-i S_0/\hbar} \langle u_0| \left[ {\hat V}_0(x) + {\hat V}_1 (-i \hbar \partial_x) + {\hat V}_2 (-i \hbar \partial_x)^2 \right] \left( {\cal A}(x) e^{i S_0/\hbar} |u_0\rangle \right) \\
	&~~~~~~~~~~~~~~=  \langle u_0|{\hat H}^s|u_0\rangle {\cal A}(x)  -i \hbar {\hat V}_{2,00} {\cal A}(x) S''_0(x) - i \hbar {\hat V}_{1,00} \partial_x {\cal A}(x) -i 2 \hbar {\hat V}_{2,00} \partial_x {\cal A}(x) S_0'(x) \\ \nonumber 
	&~~~~~~~~~~~~~~ - i \hbar {\cal A}(x) \langle u_0|  {\hat V}_{1} \partial_x |u_0\rangle - 2 i \hbar  {\cal A}(x) S_0'(x) \langle u_0|  {\hat V}_{2} \partial_x |u_0\rangle  \\
	&~~~~~~~~~~~~~~ =  \epsilon(x, p_x) {\cal A}(x)  -i \hbar {\hat V}_{2,00} {\cal A}(x) S''_0(x) - i \hbar \langle u_0| \frac{\partial {\hat H}^s}{\partial p_x} \partial_x \left[ {\cal A} |u_0\rangle \right],
	\end{align*}
	where we introduced the operator:
	\begin{align*}
	\frac{\partial {\hat H}^s}{\partial p_x} = {\hat V}_1 + 2 {\hat V}_2 p_x
	\end{align*}
	and the real (since $V_{1,2}$ are self adjoint) matrix elements ${\hat V}_{1,00} = \langle u_0| {\hat V}_{1} |u_0\rangle, \; {\hat V}_{2,00} = \langle u_0| {\hat V}_{2} |u_0\rangle$.
	
	On the other hand since $|\psi(x)\rangle$ is an eigenvector the following property takes place:
	\begin{align*}
	e^{-i S_0/\hbar} \langle u_0 | {\hat H} |\psi(x)\rangle = E {\cal A}.
	\end{align*}
	
	On semiclassiclal trajectories $\epsilon(x, p_x) = E$. Replacing the differentiation with respect to $x$ to the differentiation with respect to time $t$ we are thus led to the equation on the slowly varying complex amplitude ${\cal A}$:
	\begin{align}
	{\hat V}_{2,00} {\cal A}(x) \frac{d p_x}{dt} + \langle u_0| \frac{\partial {\hat H}^s}{\partial p_x} \frac{d}{dt} \left[ {\cal A} |u_0\rangle \right] = 0.
	\label{eq:slowA}
	\end{align}
	
	We now want to separate the equations on the phase $\phi$ and modulus $A$ of the complex quantity ${\cal A} = A e^{i \phi}$, for this purpose we rewrite Eq.~(\ref{eq:slowA}) in the form:
	\begin{align}
	{\hat V}_{2,00} {\cal A}(x) \frac{d p_x}{dt} + \frac{\partial \epsilon}{\partial p_x} \frac{d {\cal A}}{dt} + \langle u_0| \frac{\partial {\hat H}^s}{\partial p_x} \frac{d}{dt}  |u_0\rangle   {\cal A} = 0,
	\label{eq:slowA2}
	\end{align}
	where we used $\langle u_0| \frac{\partial {\hat H}^s}{\partial p_x} |u_0\rangle  = \frac{\partial \epsilon}{\partial p_x}$ (that can be obtained by differentiating over $p_x$ the identity $\langle u_0| {\hat H}^s |u_0\rangle = \epsilon$ ).
	
	In Eq.~(\ref{eq:slowA2}) the coefficient before $\frac{d {\cal A}}{dt}$ is real, what allows us to write an equation on the phase $\phi$:
	
	\begin{align*}
	{\hat V}_{2,00} A(x) \frac{d p_x}{dt} + \frac{\partial \epsilon}{\partial p_x} \left( A' + i \phi' A \right)  + \langle u_0| \frac{\partial {\hat H}^s}{\partial p_x} \frac{d}{dt}  |u_0\rangle   A  = 0.
	\end{align*}
	
	Further separating real and imaginary parts we find:
	\begin{align}
	\frac{d \phi}{dt} = -\left( \frac{\partial \epsilon}{\partial p_x} \right)^{-1} {\rm Im} \langle u_0| \frac{\partial {\hat H^s}}{\partial p_x} \frac{d}{dt} |u_0\rangle.
	\label{eq:phi1}
	\end{align}
	
	To transform this equation on the phase into the usual form, we notice the relation:
	\begin{align*}
	\frac{\partial {\hat H^s}}{\partial p_x} |u_0\rangle &= \frac{\partial}{\partial p_x} \left( {\hat H^s} |u_0\rangle \right) - {\hat H^s} \frac{\partial |u_0\rangle}{\partial p_x} \\
	&= \frac{\partial \epsilon}{\partial p_x} |u_0\rangle + (E - {\hat H^s}) |\frac{\partial u_0}{\partial p_x}\rangle.
	\end{align*}
	Since $\frac{\partial {\hat H^s}}{\partial p_x}$ is self adjoint we use this relation to transform Eq.~(\ref{eq:phi1}) into:
	\begin{align}
	\frac{d \phi}{dt} = -{\rm Im} \langle u_0| \frac{d}{dt} |u_0\rangle - \left( \frac{\partial \epsilon}{\partial p_x} \right)^{-1} {\rm Im} \langle \frac{\partial u_0}{\partial p_x}|(E - {\hat H^s}) \frac{d}{dt} |u_0\rangle .
	\label{eq:phi2}
	\end{align}
	
	Eq.~(\ref{eq:phi2}) can be simplified further by noticing that:
	\begin{align*}
	\frac{d}{dt} |u_0\rangle &= \frac{\partial |u_0\rangle}{\partial x} \frac{d x}{d t} + \frac{\partial |u_0\rangle}{\partial p_x} \frac{d p_x}{d t} \\
	&=  \frac{\partial |u_0\rangle}{\partial x} \frac{\partial \epsilon}{\partial p_x} - \frac{\partial |u_0\rangle}{\partial p_x} \frac{\partial \epsilon}{\partial x}.
	\end{align*}
	The last term gives a real contribution which does not contribute to the phase evolution, we are thus finally lead to the equation on the phase from Littlejohn~\cite{Littlejohn1991a}:
	
	\begin{align}
	\frac{d \phi}{dt} &= -{\rm Im} \langle u_0| \frac{d}{dt} |u_0\rangle - {\rm Im} \langle \frac{\partial u_0}{\partial p_x}|(E - {\hat H^s}) |\frac{\partial u_0}{\partial x}\rangle \nonumber \\
	&= i \langle u_0| \frac{d}{dt} |u_0\rangle + \frac{i}{2} \left[  \langle \frac{\partial u_0}{\partial p_x}|({\hat H^s}-E) |\frac{\partial u_0}{\partial x}\rangle - \langle \frac{\partial u_0}{\partial x}|({\hat H^s}-E) |\frac{\partial u_0}{\partial p_x}\rangle \right].
	\label{eq:phi3}
	\end{align}
	Here the first term is the well known Berry phase, while the second one does not have the name. It can be integrated along orbits just as
	well as the Berry's-phase term, but the result in general cannot be
	represented as a line integral of a differential form, thus it is not "geometrical" in the same sense as Berry's phase term and thus can be called as "non-geometric" phase. Further we will use this name to denote the second term. 
	
	\section{Appendix C: Semiclassical quantization for HgTe quantum well}
	
	Without magnetic field, the Hamiltonian of the quantum well takes the form:
	\begin{align*}
	{\hat H}_{qw} = {\hat H}_{qw}(k_x, k_y).
	\end{align*}
	In magnetic field in Landau gauge the Hamiltonian becomes:
	\begin{align*}
	{\hat H} = {\hat H}_{qw}(-i \hbar \partial_x , e B x).
	\end{align*}
	
	Rotation invariance gives the property (which can be checked by the direct calculation): 
	\begin{align*}
	& e^{-i \theta {\hat J}_z} {\hat H}_{qw}(\mathbf{k}) e^{i \theta {\hat J}_z} = {\hat H}_{qw}( R_z(\theta)(\mathbf{k}) ),
	\end{align*}
	where $J_z$ is $z$-component of the total angular momentum, $R_z(\theta)$ is the rotation operator around $z$-axis to the angle~$\theta$.
	
	We can thus write the wave function as:
	\begin{align}
	| u(k, \theta) \rangle = e^{-3i \theta/2} e^{-i \theta {\hat J}_z} | u(k, 0) \rangle,
	\label{eq:rotinv}
	\end{align}
	where the factor phase $e^{-3i \theta/2}$ is introduced to make the wavefunctions single valued upon a full rotation $\theta \rightarrow \theta + 2 \pi$ (it is necessary because $J_z$ is a direct sum of spin 1/2 and 3/2 operators) and to make $\Gamma(C)$ in range $0 \leq \Gamma(C) <  2\pi$.
	
	The Berry phase $\Gamma_B$ (the first term in Eq.~(\ref{eq:phi3})), is thus simply connected with the spin of the wavefunctions:
	\begin{align*}
	\Gamma_B &= i \int d \theta \langle  u(k, \theta) | \frac{d}{d \theta} | u(k, \theta) \rangle  \\
	&=  \int d \theta \langle  u_0 | {\hat J}_z + \frac{3}{2} | u_0 \rangle \\
	&= 2 \pi \langle J_z\rangle + 3\pi,
	\end{align*}
	where we introduced	$|u_0\rangle = |u(k, 0)\rangle$.
	
	The non-geometric contribution (the second term in Eq.~(\ref{eq:phi3})) can be also simplified taking into account the rotation invariance~(\ref{eq:rotinv}):
	\begin{align*}
	\Gamma_{NG} = \int dt \gamma_{NG}.
	\end{align*}
	
	We first write $\gamma_{NG}$ in cylindrical coordinates:
	\begin{align*}
	\gamma_{NG} &= \frac{i e B}{2} \left[ \langle \frac{\partial u}{\partial p_y}|({\hat H^s}-E) |\frac{\partial u}{\partial p_x}\rangle - \langle \frac{\partial u}{\partial p_x}|({\hat H^s}-E) |\frac{\partial u}{\partial p_y}\rangle \right] \\
	&= \frac{i e B}{2} \left[ \langle \frac{\partial u}{\partial p} \frac{\partial p}{\partial p_x} + \frac{\partial u}{\partial \theta} \frac{\partial \theta}{\partial p_x}|({\hat H^s}-E) | \frac{\partial u}{\partial p} \frac{\partial p}{\partial p_y} + \frac{\partial u}{\partial \theta} \frac{\partial \theta}{\partial p_y} \rangle - \langle \frac{\partial u}{\partial p_x}|({\hat H^s}-E) |\frac{\partial u}{\partial p_y}\rangle \right] \\
	&= \frac{i e B}{2} \left( \frac{\partial \theta}{\partial p_x} \frac{\partial p}{\partial p_y} - \frac{\partial \theta}{\partial p_y} \frac{\partial p}{\partial p_x} \right) \left[  \langle \frac{\partial u}{\partial \theta} |({\hat H^s}-E) | \frac{\partial u}{\partial p} \rangle - \langle \frac{\partial u}{\partial p}|({\hat H^s}-E) |\frac{\partial u}{\partial \theta}\rangle \right] \\
	&= -\frac{i e B}{2 p} \left[  \langle \frac{\partial u}{\partial \theta} |({\hat H^s}-E) | \frac{\partial u}{\partial p} \rangle - \langle \frac{\partial u}{\partial p}|({\hat H^s}-E) |\frac{\partial u}{\partial \theta}\rangle \right],
	\end{align*}
	where $p = \sqrt{p_x^2 + p_y^2}$.
	And then we use (\ref{eq:rotinv}) to obtain:
	\begin{align*}
	\gamma_{NG} =  \frac{e B}{2 p} \left[ \langle u_0 | \left( {\hat J}_z + \frac{1}{2} \right) ( {\hat H^s}-E ) | \frac{\partial u_0}{\partial p} \rangle +  \langle \frac{\partial u_0}{\partial p} |  ( {\hat H^s}-E ) \left( {\hat J}_z + \frac{1}{2} \right) |u_0\rangle  \right],
	\end{align*}
	where the fact that ${\hat J}_z$,  $e^{-i \theta {\hat J}_z}$ and  ${\hat H^s}$ all commute with each other was used. We thus find that $\gamma_{NG}$ does not depend on $\theta$ but only on the momentum amplitude. 
	
	Since $\gamma_{NG}$ is constant along the semiclassical trajectory the total non-geometric phase is:
	\begin{align*}
	\Gamma_{NG} = \frac{2 \pi}{\omega_c} \gamma_{NG},
	\end{align*}
	where $\omega_c = \frac{e B}{p} \frac{d \epsilon}{d p}$ is the (semiclassical) cyclotron frequency at energy $E = \epsilon(p)$. 
	
	Finally we find:
	\begin{align*}
	\Gamma_{NG} &=  2 \pi \; {\rm Re}  \langle u_0 | \left( {\hat J}_z + \frac{1}{2} \right) ( {\hat H^s}-E ) | \frac{\partial u_0}{\partial \epsilon} \rangle \\
	&= 2 \pi \; {\rm Im}  \langle \frac{\partial u}{\partial \theta} | {\hat H^s}-E | \frac{\partial u}{\partial \epsilon} \rangle.
	\end{align*}
	
	\clearpage
	\section{Appendix D: Well assymetry due to gate potential}
	
	\begin{figure}[H]
		\begin{minipage}[h]{1\linewidth}
			\begin{tabular}{p{0.5\linewidth}p{0.5\linewidth}}
				(a) & (b) \\
			\end{tabular}
		\end{minipage}
		\begin{minipage}[h]{0.5\linewidth}
			\flushright\includegraphics[scale=1.15]{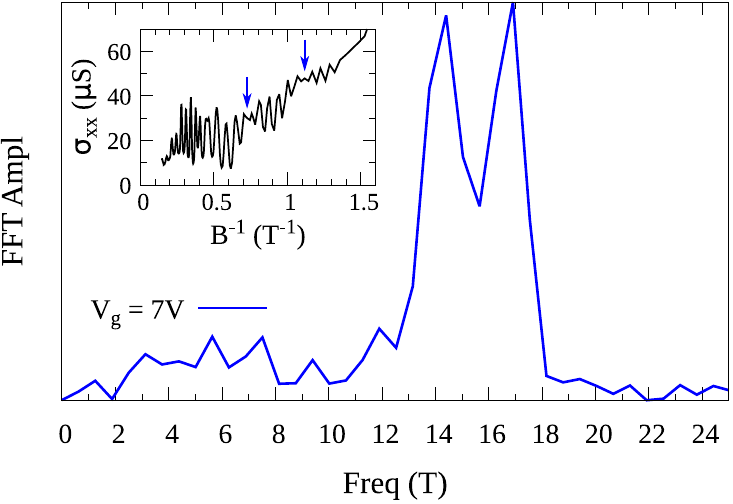}
		\end{minipage}
		\begin{minipage}[h]{0.45\linewidth}
			\flushright\includegraphics[scale=1.15]{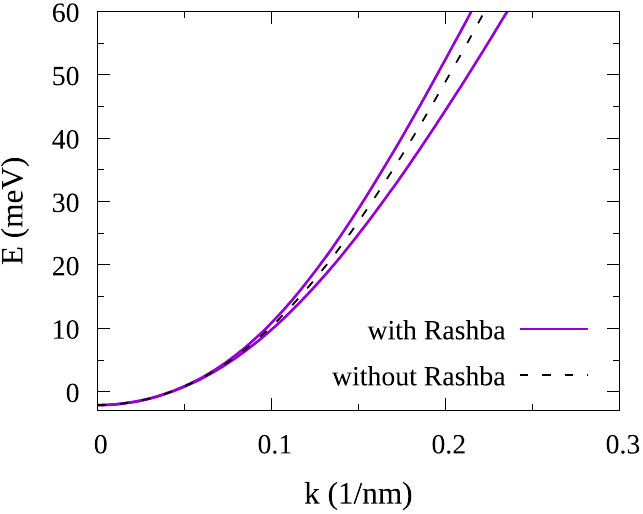} \\
		\end{minipage}	
		\vfill
		\begin{minipage}[h]{1\linewidth}
			\begin{tabular}{p{0.5\linewidth}p{0.5\linewidth}}
				(c) & (d) \\
			\end{tabular}
		\end{minipage}	
		\begin{minipage}[h]{0.5\linewidth}
			\flushright\includegraphics[scale=1.15]{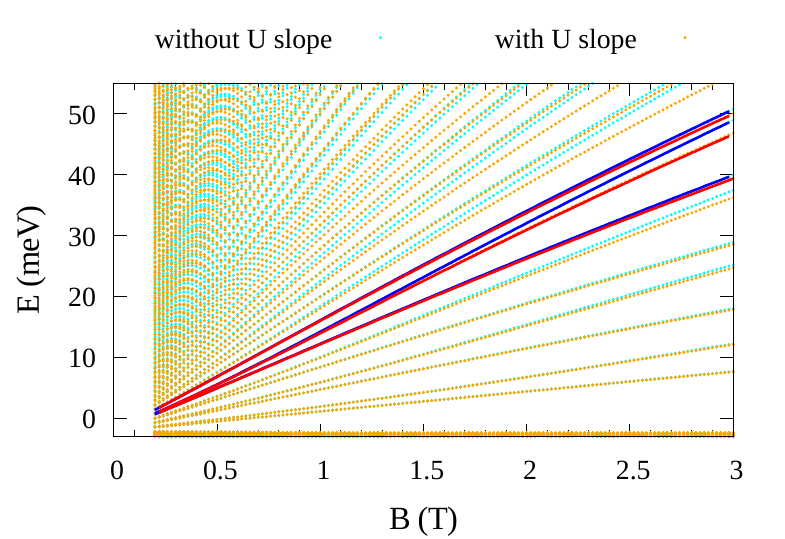} \\
		\end{minipage}
		\begin{minipage}[h]{0.5\linewidth}
			\center\includegraphics[scale=1.15]{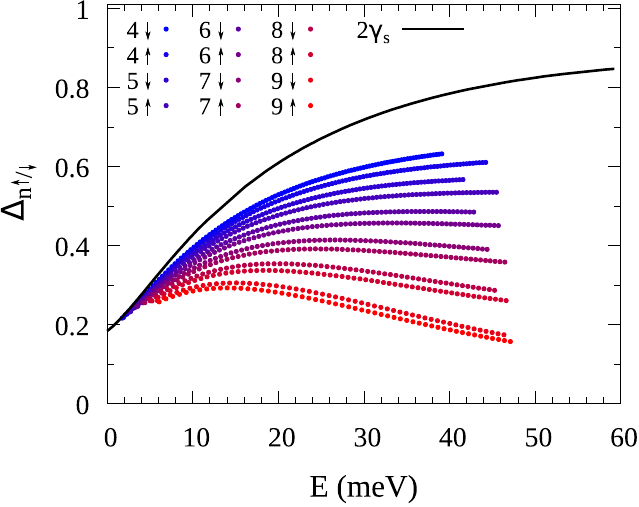} \\
		\end{minipage}
		\caption{(a) Fourier spectrum of longitudinal conductivity at gate voltage of 7\,V. Due to Rashba splitting two peaks at frequencies 14.1 and 16.8\,T are seen. \textit{Inset}: longitudinal conductivity versus inverse magnetic field. Arrows indicate beatings due to Rashba splitting. (b-d): Calculations for (001) 20\,nm HgTe strained quantum well by isotropic 6-band Kane model. (b)~Energy spectrum of the conduction band without (black dashed line) and with (violet lines) taking into account well deformation arising as a consequence of the applied top gate voltage. The gate voltage effect was modeled as a linear well bottom slope growing linearly from zero at conduction band minimum to 25\,mV ( electric field 12.5\,kV/cm) at energy 60\,meV. (c)~Landau levels position as the function of the perpendicular magnetic field calculated with (orange dots) and without (cyan dots) taking into account the well asymmetry due to gate voltage. The blue lines show the Landau levels with non-trivial behavior that were highlighted in Fig.~3 in the main text. The red lines correspond to the same Landau levels, but calculated with considering the well asymmetry due to gate voltage. The qualitative behavior of red lines including well anisotropy is similar to the blue ones, but a bit further from experimental Landau levels behavior. Panel (d) displays the ratio between Zeeman and orbital level splittings $\Delta_{n \uparrow / \downarrow}$ (dots) as a function of electron energy taking into account the well asymmetry induced by the perpendicular electric field. Since the conduction band is spin-split (see Panel b) the universal scaling that observed in Fig.~4\,(a) from the main text is no longer strictly followed. }
		\label{pic:LLs}
	\end{figure}
	
	The electron density in the quantum wells was changed applying a top gate voltage. It is known that this results in an asymmetrical well deformation, which leads to the formation of spin-polarized bands due to the Rashba effect, which in turn leads to a shift of the Landau levels ~\cite{Bychkov1984,Zhang2001}. Here we show that the Rashba effect does not change much the qualitative picture of Landau levels behavior, which was discussed in the main text, and does not help to improve the quantitative agreement with the experiment. A rigorous approach to treat the gate voltage effect requires solving self-consistent Poisson and Schr\"odinger equations, in our case we used the simple approximation and modeled the effect of the top gate as a linear potential gradient inside the well (in other words a constant electric field). The gradient is zero at energies corresponding to the bottom of the conduction band and increases linearly with electron energy to 25\,mV at energy $E=60$\,meV corresponding to electric field of 12.5\,kV/cm. The slope has been chosen in the way that calculated Rashba splitting at energy 56.3\,meV (corresponds to gate voltage of 7\,V) was the same as obtained experimentally from Fourier analysis of Shubnikov - de Haas oscillations (see Fig.~\ref{pic:LLs}\,(a)). The energy spectrum of the conduction band calculated with and without taking into account the well asymmetry due to top gate voltage is shown in~Fig.~\ref{pic:LLs}\,(b).
	
	The comparison of the numerical results for Landau levels of the first conduction band versus magnetic field with (cyan dots) and without (orange dots) taking into account the Rashba effect is shown in Fig.~\ref{pic:LLs}\,(c). An ensemble of three Landau levels highlighted by blue lines shows the non-trivial behavior where Landau levels split and come closer to each other again when magnetic field increases, it corresponds to the energy levels highlighted by lines in Fig.~3 in the main text. Red lines correspond to the same Landau levels, but calculated with considering the Rashba spin-splitting. The qualitative behavior of ''red'' Landau levels is close to blue lines, however at high magnetic field two upper red lines are more separate than the blue ones. Rashba effect thus does not enhance the intrinsic effect due to the mixing of the conduction band with the second well subband, but makes it a bit weaker.
	
	Also as it can be seen from Fig~\ref{pic:LLs}\,(d) the ``universal'' scaling from Fig.~4\,(a) of the main text, where the ratio between Zeeman and orbital Landau level splitting collapsed on a single curve given by the doubled electronic phase $2 \gamma_s$, is no longer valid in presene of Rashba interaction. This occurs because the derivation of the scaling relation assumed the same band structure for both spin orientations which is no longer correct when Rashba interaction is present. We see that due to Rashba interaction the dependence of $\Delta_{n \uparrow / \downarrow}$ on energy becomes weaker with Landau level number increase, which is at odds with the experimental results (see Fig.~4\,(b) in the main text) where this dependence becomes instead stronger at higher Landau level numbers.
	
	To conclude this part, we have shown that Rashba interaction does not change the qualitative picture presented in the main manuscript, quantitatively the induced correction seem to show the opposite trend compared to the experimental results suggesting that other physical effects need to be included. 
	
	\clearpage
	\section{Appendix E: Comparison of non-isotropic and isotropic Landau levels calculations}
	
\begin{figure}[H]
		\centering
		\includegraphics[scale = 2]{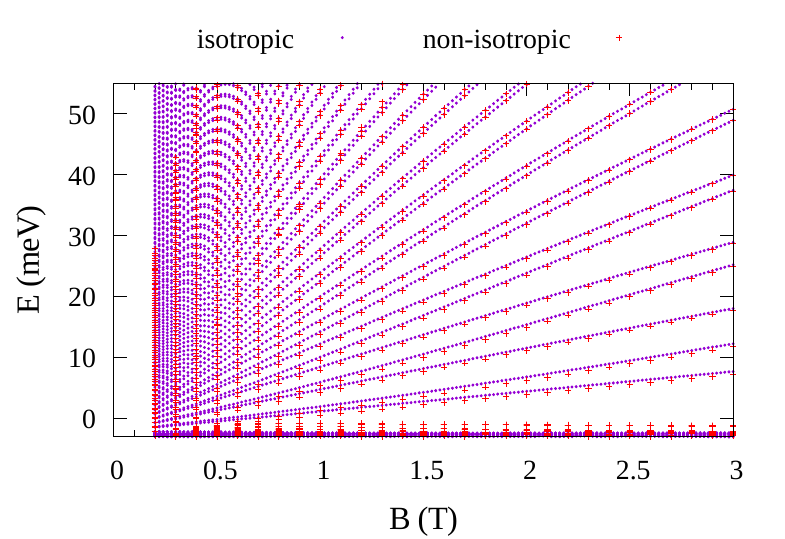}
		\caption{Landau levels position as a function of the perpendicular magnetic field in (001) 20\,nm HgTe quantum well calculated by 6-band Kane model with (violet points) and without (red crosses) isotropic approximation. The deviation of the calculation with isotropic approximation from it in non-isotropic case is seen being small and decrease at high Landau level numbers.}
		\label{pic:LLsFull}
	\end{figure}

	In order to calculate the position of Landau levels as a function of the perpendicular magnetic field, we used an isotropic model for the quantum well. This assumption enabled us to find eigenstates using only a small number of nearby oscillator wavefunctions and thus to solve the problem without diagonalizing very large matrices. In Fig.~\ref{pic:LLsFull} the comparison of the calculations with (violet points) and without (red crosses) isotropic approximation is presented. In non-isotropic case 40 Landau levels were included in the calculation what made the diagonalizing matrix size being 24\,000. The deviation of the calculation with isotropic approximation from it in non-isotropic case is seen being small and decrease at high Landau level numbers.
	
\end{document}

%% file: ShdHO_arXiv_bbl.bbl
%